\documentclass[aps,apsfig,pre,twocolumn]{revtex4}
\usepackage{graphicx}

\begin{document}
\title{Stabilization of ratchet dynamics by weak periodic signals}
\author{Maria Barbi}\thanks{e-mail address: {\tt barbi@sa.infn.it}}
\author{Mario Salerno}\thanks{temporarily at the
Physics Department of DTU, Denmark, DK-2800, Lyngby, Denmark;
e-mail: {\tt salerno@sa.infn.it}}
\address{Dipartimento di Fisica "E.R.Caianiello",\\
  Universit\`a di Salerno, Italy\\
  and \\
  Istituto Nazionale di Fisica della Materia (INFM),
  Unit\`a di Salerno, Italy. }

\begin{abstract}
  We study the influence of weak periodic signals on the transport
  properties of underdamped ratchets.  We find that the constant
  current intervals related to the ratchet, can be significantly
  enlarged by a weak subharmonic signal which is in phase with the
  internal driver.  This stabilization phenomenon is found to exist
  both in absence and in presence of noise.  The dependence of this
  effect on the phase of the applied signal is also investigated.
\end{abstract}

\maketitle

\section{Introduction}

\label{intro} 

Ratchet systems, i.e. Brownian particles moving in asymmetric periodic
potentials, have been largely investigated in the last ten years,
since their first version was introduced by Ajdari and Prost
\cite{Ajd92}.  These systems are maintained far from equilibrium by
periodic or correlated stochastic forces which can be either
multiplicative or additive, (the corresponding models being called
{\em flushing-potential} and {\em fluctuating-force} ratchets,
respectively), so that the thermal bath energy can be converted into
effective work without any conflict with the second law of
thermodynamics \cite{Feyn}. The damping and the asymmetry of the
potential are crucial ingredients for this conversion, both in the
multiplicative \cite{Ajd92,Doe95,Ast94} and in the additive
\cite{Ast94,Mag93,Doe98,Han96} case. This phenomenon arises in a
variety of different systems and has been used to design new
experimental devices both for physical and biological applications
\cite{Fau95,Gor98,Ket00,Gor96,Ert98,Gor97,Bad99}.  Moreover, the ratchet
effect is presently considered as a possible mechanisms by which
molecular motors (e.g. kinesins, myosins, dyneins) take advantage of
thermal fluctuations to perform their functions
\cite{Ast94,Mag94,Cor92,Ast99,Jul97,Kel00}.

On the other hand, {\it fluctuating-force} ratchet dynamics are
possible also in absence of noise, both in overdamped systems
\cite{Cha95,Han96,Dia97} and in underdamped chaotic ones
\cite{Jun96,Mateos}. In a previous paper \cite{Barbi} the ratchet
motion of a particle subject to an additive periodic forcing ({\it
  fluctuating-force ratchet}) was ascribed to phase locking between
the motion of the particle in the asymmetric potential and the
frequency of the driver.  The current steps arising from this phase
locking dynamics were well preserved (at least the relevant ones) also
in presence of noise, with a tendency of decreasing in width as the
noise intensity was increased.  Thus, at least for these types of
ratchets, phase locking emerges as the basic mechanism underlying
ratchet dynamics, both in presence and in absence of noise.

Since phase locking is a very well known and largely investigated
phenomenon in many areas of science, one can take advantage of its
knowledge to infer results in the field of ratchets.  Thus, for
example, in Refs. \cite{Sal91,Sal94} it was shown that the phase
locking steps arising in the voltage-current characteristic of a long
Josephson junction can be stabilized by applying weak subharmonic
signals which suppress the deterministic chaos \cite{comment}.  The
above phase locking interpretation of the ratchet dynamics naturally
suggests that similar stabilization phenomena could exists also in
{\it fluctuating-force} ratchet models.

The present paper is just devoted to this, i.e.  we study the effects
of weak subharmonic signals on the ratchet dynamics of an underdamped
particle moving in an asymmetric potential both in absence and in
presence of noise. The aim of the paper is twofold.  From one side we
are interested to enlarge the regions of the parameter space for which
stable direct currents are observed.  We find that this is indeed
possible and better achieved when weak subharmonic signals, in-phase
with the internal driver, are applied.  The stabilization effect is
observed both in presence and in absence of noise and is accompanied,
in analogy with Josephson junctions \cite{Sal91}, by a suppression of
the deterministic chaos, a property which can be useful in technical
applications.

On the other side, we are interested in {an external control on} the
functioning of the ratchet mechanism.  We find that, depending on the
relative phase between external and internal drivers, one can
stabilize different orbits of the system, as well as, destabilize the
ratchet.  Thus, when the unidirectional motion of a physical or
biological system is governed by an additive ratchet, it could be
possible to control its dynamics by applying suitable "out of phase"
subharmonic signals.

In many practical situations, however, the relative phase between the
internal and the external (subharmonic) driver could be difficult to
control. In these cases the phase should be considered as a random
variable and a final average on it should be performed in the
calculation of the current.  We find that also in this case, although
reduced, the subharmonic signal induces a stabilization on the ratchet
dynamics.

The paper is organized as follows. In Section \ref{first} we introduce
the model and discuss the stabilization induced by a weak subharmonic
field, in-phase with the internal driver, both in absence and in
presence of noise.  In Section \ref{second} we investigate the effects
of a relative phase between the two drivers on the stabilization
phenomenon. Finally, the main results of the paper will be resumed in
the conclusions.  \newline

\section{Subharmonic stabilization effects}

\label{first} 
To conform with previous studies we take the same model as in Ref.
\cite{Barbi}, i.e. we consider a particle that moves in the spatially
periodic asymmetric potential ${\cal V} (x)$
\begin{equation}  \label{potential}
{\cal V}(x) = C - \frac{1}{4 \pi^2 \delta} \left[ \sin{(2 \pi (x-x_0))} + 
\frac{1}{4} \sin{(4 \pi (x-x_0))} \right] ,
\end{equation}
subjected to time-periodic forcing, damping, and noise.  In
Eq.~(\ref{potential}) $C$ and $x_0$ are introduced in order to have
one potential minimum in $x=0$ with ${\cal V}(0)=0$, and $\delta =
\sin{(2 \pi|x_0|)} + \frac{1}{4} \sin{(4 \pi |x_0|)}$.  Since we are
interested in stabilization effects, we introduce also a small
subharmonic signal, so that the equation of motion, in dimensionless
variables, is
\begin{equation}  \label{sub}
\ddot{x}+b \dot{x}+\frac{d{\cal V}(x)}{dt}+2D \xi(t) 
= a \cos{(\omega t+\phi)} + c \cos{(\frac{\omega}{2}\,t)} \,.
\end{equation}
Here $b$ is the friction coefficient, $\xi(t)$ is a white noise
fluctuation and $D$ its intensity, $\omega$, $a$ are respectively the
frequency and amplitude of the internal driver, $c$ is the amplitude
of a small ($c\ll 1$) subharmonic field, and $\phi$ a relative phase.


In the deterministic case and in absence of the subharmonic signal
($D=0\,,c=0$), it is known that net average motion in one direction
arises when the time required for the particle to move from one well
of the potential to another is commensurable with the period of the
internal driver, i.e. when the particle motion becomes locked to the
driver \cite{Barbi}.  The mean velocity of the particle stays constant
for all parameter values for which the locked solution is stable
(locking range), and is given by
\begin{equation}  \label{rule}
\langle v \rangle_t\, = \frac{m}{n} \, \frac{L}{T} = \frac{m}{n} \, \frac{
\omega}{2 \pi} \, L = \frac{m}{n} \, V\,,\hspace{6mm} n \in N, m \in Z \,
\end{equation}
where $L$ is the spatial period of the potential (in our case $L=1$),
$T=2 \pi / \omega$ and we call $V$ the fundamental locked mean
velocity induced by the driver $\omega L / 2 \pi$ (the current is
calculated as the particle velocity averaged over time, or, briefly,
{\em mean velocity}, $\langle v \rangle_t$). When the conditions of
Eq.~(\ref{rule}) are achieved the particle follows regular orbits in
the phase space, otherwise it displays a chaotic motion with zero mean
velocity \cite{Barbi}.  To simplify our study we fix in all the
following numerical simulations, $\omega=0.67$ and $b=0.1$
(qualitatively similar results are obtained for other parameter
values) and consider $a$ and $c$ to be free parameters.

The solid curve reported in Fig.~1 represents the average velocity
(current) of the particle vs the internal driver amplitude for $c=0$
and in absence of noise (the average is taken over $300$ forcing
periods, with a time step of $T/1000$). Current steps with velocities
$0$, $V/2$, $-V/4$, $-V/2$, as well as chaotic regions without locking
effects, are clearly recognized.  To investigate stabilization
phenomena induced by the subharmonic driver we consider $c \ne 0$ in
Eq.~(\ref{sub}) and focus, for simplicity, on the largest current step
in Fig.~1 (solid curve) corresponding approximately to the range
$a\in(0.062,0.076)$ \cite{hyst}.


Let us start first with the case of zero noise and zero relative phase
between the two drivers. The stabilization effect in this case is seen
from the enlargement of the steps computed for $c=0.001$, $c=0.002$,
$c=0.003$ and $c=0.004$ as reported by the broken curves of Fig.~1
(the curves with increasing values of $c$ were vertically shifted to
avoid overlapping).  In the inset of the figure the step width as a
function of $c$ is also reported, from which we see that the width
increases roughly linearly with the subharmonic amplitude. As expected
from Ref.\cite{Sal91}, the subharmonic signal tends to suppress the
chaos present in the system, and at $c=0.004$ the chaotic region near
$a=0.078$, visible in the $c=0$ case, disappears completely.  At
higher values of $c$, the situation becomes more involved (steps can
``break'' and instabilities arise) due to the more complicated
structure of the phase space of the two-drivers system (in these cases
the subharmonic is not anymore a small perturbation).

Let us now introduce noise in the system but still with the relative
phase $\phi$ fixed to zero. We choose a noise intensity $D=10^{-6}$,
corresponding, in the dimensional parameter space, to approximatively
thermal noise at room temperature for a system with mass
$m\sim200\,k\,a.m.u.\sim3.3\,10^{-22} kg$, spacing $L\sim 8 \, nm$ and
a potential barrier of $8\, kT$ \cite{Barbi}.


In Figure~2 we present the results of the application of the same
external driving forces considered in the deterministic case to a
population of $50$ particles in presence of noise.  We have introduced
again a vertical shift for each curve for the aim of readability of
the picture (the mean velocity is now averaged also over all
particles, $\langle\langle v \rangle_t \rangle_N$). At $D=10^{-6}$ and
$c=0$ (Figure~2, solid line), almost all the steps in the same range
of Figure~1 disappear; nevertheless, the largest one is preserved and
still corresponds to a constant velocity $V/2$. Note however how its
width is reduced in presence of noise.  Even in this case, the
stabilization effect of the subharmonic driving term results with
evidence. The step in current is more and more enlarged by the
application of the external forcing with increasing amplitudes (the
step width as a function of $c$ is displayed in the inset of Fig.~2).

From these results we see that the region of the internal parameter
space for which a stable direct current is observed can be
significantly enlarged by the application of a weak subharmonic field,
in-phase with the internal driver, either in absence or in presence of
noise. Moreover, we observe that the stabilization effect is always
associated with a suppression of the deterministic chaos.

\section{Phase dependent stabilization}

\label{second} 



In natural systems, such as biological motors, the phase of the
internal driver is an unknown parameter so that $\phi$ becomes
difficult (if not impossible) to control.  Furthermore, a population
of many of such motors should correspond to a set of model particles
with random phases, their internal drivers being, in principle, not
synchronized.  It is therefore interesting to study the effect of the
phase $\phi$ on the stabilization. Since the noise leads to a
smoothing of the current steps with a reduction of their widths, we
shall concentrate only on the deterministic $D=0$ case.

To this end we have studied the effect of the phase $\phi$ for some
fixed values of the internal and external driver amplitude.  The
external driver was fixed to $c=0.004$ in all cases.  Letting evolve
the system according to Eq.~(\ref{sub}) for each different value of
$\phi$ in $(0,\, 2 \pi)$, we obtained for different values of $a$ the
mean velocity as a function of $\phi$.  Results for some interesting
values of $a$ are shown in Figure~3.  The first three value of the
internal driver of Figure~3, i.e.  $a=0.077$, $0.078$, $0.079$, all
correspond to points on the current step (see Figure~1, dotted line).
The effect of the subharmonic driver results to be indeed phase
dependent. While the mean velocity at $\phi=0$ is $V/2$ in all three
cases, the stabilization of the corresponding orbit arises only in
some ranges of the parameter $\phi$.  When the internal and external
drivers are approximatively around phase $\pi/ 2$ or $3 \pi / 2$, it
can happens that the subharmonic tends to destabilize the system to a
chaotic orbit, as for $a=0.077$; in other cases, it tends instead to
stabilize a symmetrical orbit with opposite velocity $-V/2$, as for
$a=0.078$.  The stabilization regions, however, are dominant with
respect to the chaotic ones. This is a consequence of the coexistence
of different regular orbits at a time \cite{Barbi} so that when one
orbit destabilizes, another orbit (with different current) becomes
available for transport (the chaos is only in the transient from one
stable motion to another). This is very similar to what described in
Ref. \cite{Sal94} on the stabilization of the phase locking dynamics
of long Josephson junctions.

For other values of the parameter range, many different regular orbits
can be stabilized for different values of the phase, as in the case of
$a=0.079$. Note that in this case the particle motion is regular
almost everywhere.  Nevertheless, an average of the mean velocity over
all different values of the phase for values as $a=0.078$ and
$a=0.079$ will give almost zero, because of the mixing up of orbits
with opposite velocities. This has important consequences in the case
of an ensemble of particles with random phases, as discussed below.
The last case shown in Figure~3, $a=0.097$, corresponds to the second
end of the whole interval considered in Figure~1 (this region is
chaotic for $c=0$). Interestingly, while in the in-phase case the
subharmonic driver has no effects on this chaos, for some values of
the phase (again near $\phi=\pi/2$ and $3\pi/2$) the subharmonic field
can induce the stabilization of regular orbits with mean velocities
$-V$ and $-V/4$. From this we can conclude that a weak subharmonic
perturbation {\em with an appropriate phase} with respect to the
internal driver can induce phase locking in parameter regions where
only chaotic motion is present for the unperturbed system. The
application of an external subharmonic driver has therefore relevant
phase dependent effects on the ratchet dynamics.


In order to consider the phase averaged effect of a weak subharmonic
signal on a set of particles, we have performed the same analysis of
Figure~2 but introducing a random phase $\phi_i$ for each particle
$i=1, N$. We used in this case a population of $200$ particles,
$c=0.004$, and again $D=0$. Results are shown in Figure~4.  The
results obtained in the cases $\phi_i=0, \, c=0$ and $\phi_i=0, \,
c=0.004$ are also shown for comparison. According to Figure~4, the
phase averaging causes a reduction of the mean velocity in the main
locked window, and the plateau is in this case poorly defined. Note
that although the width of the step is reduced with respect to the
corresponding $\phi=0$ case (dots), it is still slightly larger that
the one in absence of subharmonic (open squares).

We also remark that the averaged mean velocity shown in Figure~4 could
lead to misleading conclusions. Indeed, the values where the averaged
mean velocity drops to zero seems coincide with the chaotic region of
the unperturbed system around $a=0.078$.  One could then be tempted to
conclude that the chaotic orbits are actually preserved in average,
with no relevant stabilization effect. This, however, is not the case
as one can see by calculating in the same range of parameters the
average of the mean velocity {\em absolute value}, $\langle |\langle v
\rangle_t| \rangle_N$.  The result is shown in the same Fig.~4, as a
dashed line. The absolute value of the velocity is clearly well above
zero for all values of $a$, and we can conclude that the zero average
of the velocity is due to the mixing of orbits with positive and
negative velocities (and not to the presence of chaotic orbits with
zero mean velocity). We finally remark that the average on the phase
leads to a lowering of the mean velocity, with a deviation of the
current step from a straight segment, this being a consequence of the
mixing of orbits with different velocities stabilized by different
phases.

\section{Conclusions}

\label{concl} 

In this work we addressed the study of the stabilization effect of a
weak subharmonic field on the phase locked dynamics of a ratchet
system.  We found that, for fluctuating-force ratchets, the
application of an external subharmonic driver suppresses chaos and
stabilizes regular orbits over larger ranges of the internal driver
amplitude.  This phenomenon strongly depend on the relative phase of
the internal and external drivers which can then be used as a control
parameter in the stabilization of a particular ratchet motion.

It would be interesting to apply these ideas to real experimental
devices such as, for example, ratchet particle separators
\cite{Ket00,Gor96}.

\section*{Acknowledgments}
M.B. thanks the EC and the University of Salerno for Post-Doc
research support.
M.S. wish to thank the Department of Physics of the Technical University
of Denmark for the hospitality and for providing
a two months Visiting Professorship during which
this work was finished.
Partial financial support from INFM and
from the European grant LOCNET
(contract number HPRN-CT-1999-00163), is also
acknowledged.

\newpage

\section*{FIGURE CAPTION}

\begin{itemize}
\item[{\bf FIG. 1}]{Mean velocity as a function of the internal drive
    amplitude $a$ for the system \ref{sub} in absence of noise ($D=0$)
    and for $\phi=0$. The solid line refers to the case $c=0$ where
    there is no subharmonic signal. The others curves correspond to
    the cases $c=0.001$, $0.002$, $0.003$, $0.004$: we have shifted
    each of them vertically in order to distinguish them more easily.
    Note how the application of subharmonics with increasing
    amplitudes tends to enlarge the main step in current, suppressing
    the chaotic region around $a=0.080$. {\sl Inset:} Width of the
    main step in the displayed curves as a function of the subharmonic
    amplitude $c$. Note that the stable region increases roughly
    linearly. All plotted variables are in dimensionless units.}
\item[{\bf FIG. 2}]{Mean velocity, averaged on a set of $50$
    particles, as a function of the internal drive amplitude $a$ for
    the system \ref{sub} in presence of noise ($D=10^{-6}$) and for
    $\phi=0$. The solid line refers to the case $c=0$ . The others
    curves correspond to the cases $c=0.001$, $0.002$, $0.003$,
    $0.004$: we have shifted each of them vertically in order to
    distinguish them more easily.  {\sl Inset:} Width of the main step
    in the displayed curves as a function of the subharmonic amplitude
    $c$. All plotted variables are in dimensionless units.}
\item[{\bf FIG. 3}]{Mean velocity as a function of the phase $\phi$
    for four different values of the internal driver ($a=0.077$,
    $0.078$, $0.079$ and $0.097$) with $c=0.04$ and $D=0$. Plotted
    variables are in dimensionless units.}
\item[{\bf FIG. 4}]{Mean velocity, averages on a set of $200$
    particles with random phases $\phi_i$, as a function of $a$, for
    $c=0.004$ (solid line). The $\phi=0$ mean velocity for the cases
    $c=0$ (open squares) and $C=0.004$ (small dots) are shown for
    comparison. The dashed line is the average over the particles of
    the absolute value of the mean velocity, $\langle |\langle v
    \rangle_t| \rangle_N$. Plotted variables are in dimensionless
    units.}
\end{itemize} 


\begin{thebibliography}{10}
\bibitem{hyst} {As shown in Ref. \cite{Barbi}, the presence of
    coexisting orbits in some ranges of the parameter $a$ could lead
    to hysteresis, so that the width of a locked step can depend on
    how $a$ is varied during the simulations. In order to reduce this
    effect, we perform our simulations in such a way that $a$ is
    always varied starting from one value inside this region and going
    first toward smaller and then toward bigger values. In this way,
    the width of the step is maximized.}
\end{thebibliography}
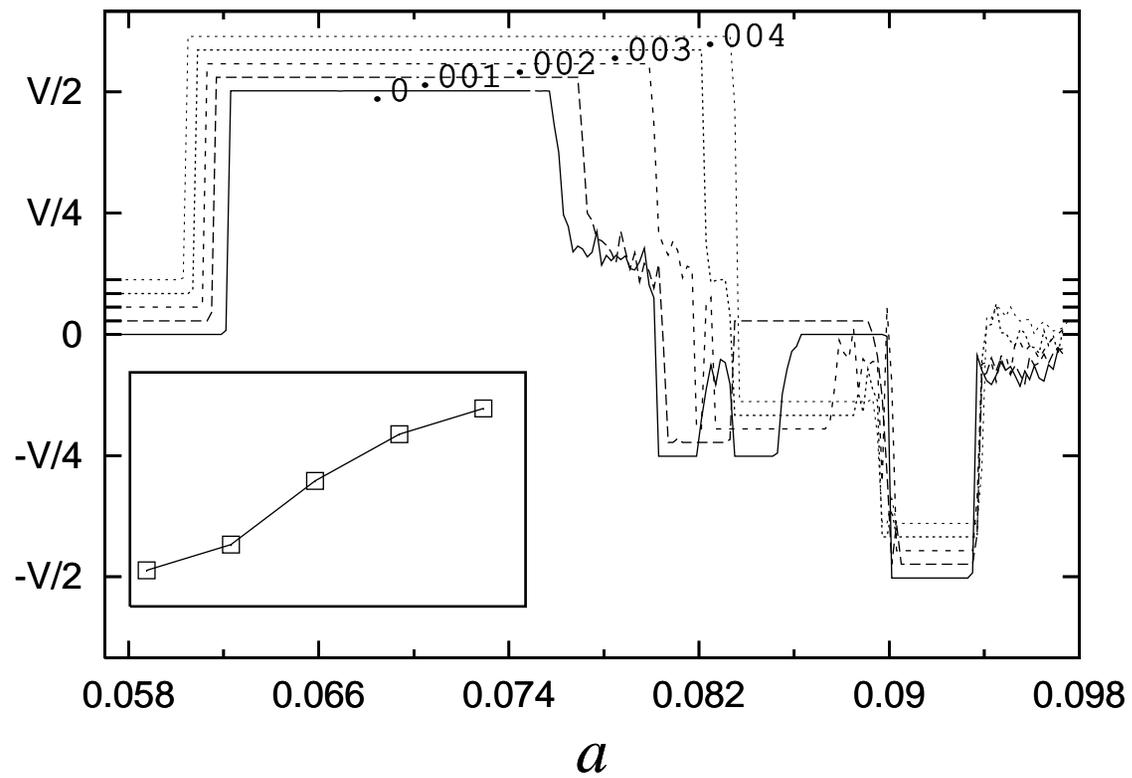
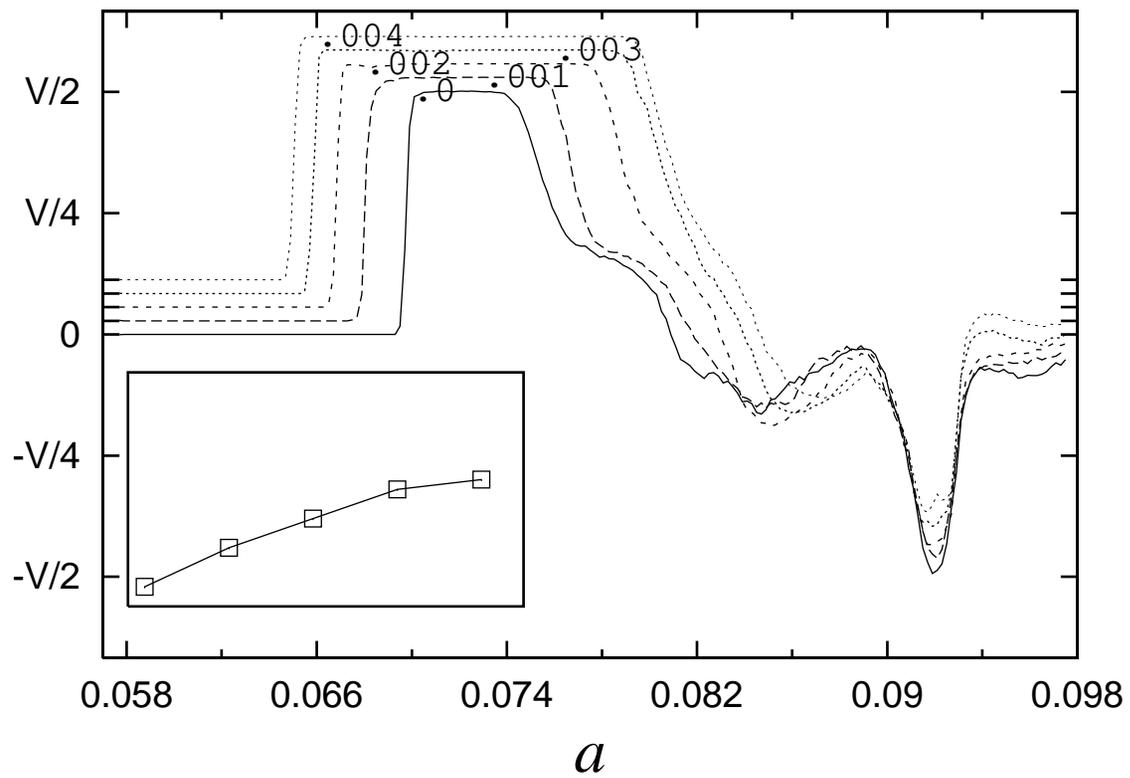
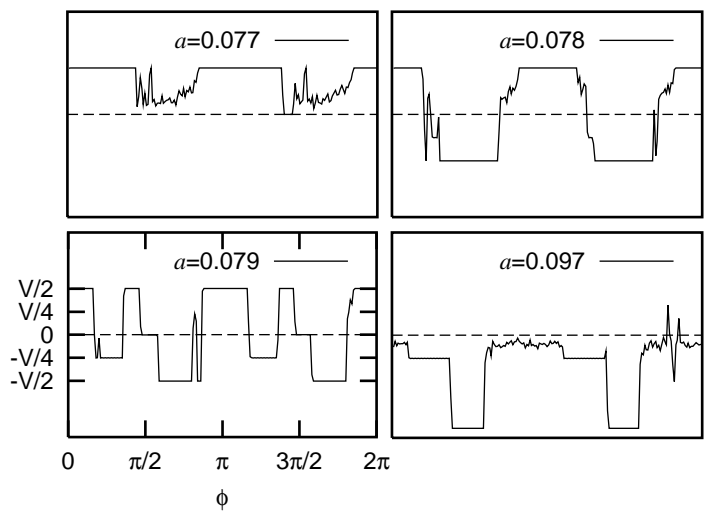
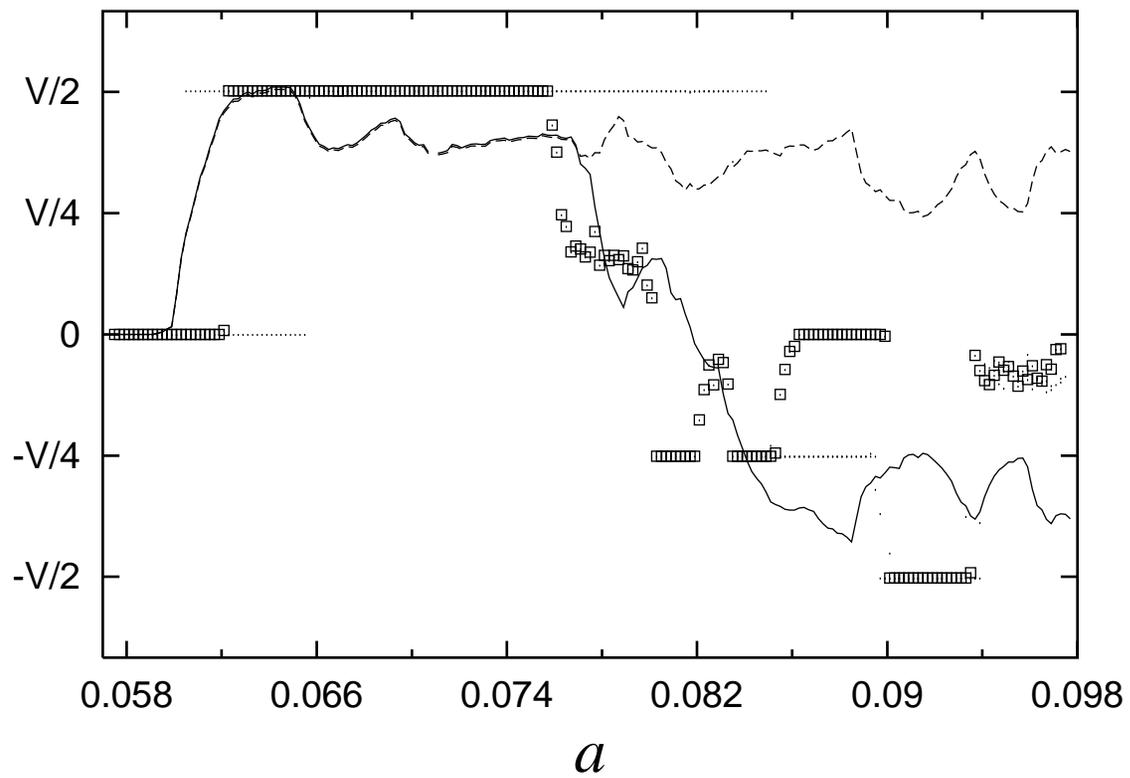
\end{document}